\documentclass[runningheads]{llncs}

\usepackage{amsmath,amssymb,amsfonts}
\usepackage[ruled,vlined,linesnumbered]{algorithm2e}
\usepackage{graphicx}
\usepackage{textcomp}
\usepackage[table]{xcolor}
\usepackage{hyperref}
\usepackage{xurl}
\usepackage{tikz}
\usepackage{mathtools}
\usepackage{booktabs}
\usepackage{multirow}
\usepackage{enumitem}
\usepackage{float}
\newcommand{\cref}[1]{\ref{#1}}
\newcommand{\Cref}[1]{\ref{#1}}

\usetikzlibrary{trees, positioning, shapes}
\usetikzlibrary{
    arrows,
    arrows.meta,
    calc,
    decorations.pathreplacing,
    shapes.geometric}

\tikzstyle{process}     = [rectangle, minimum width=3cm, minimum height=1cm,
                           text centered, draw=black, fill=orange!30]
\tikzstyle{decision}    = [diamond, minimum width=3cm, minimum height=1cm,
                           inner sep=1pt, text centered, aspect=3,
                           draw=black, fill=green!30]
\tikzstyle{startstop}   = [rectangle, rounded corners, minimum width=1.4cm,
                           minimum height=0.7cm, text centered, draw=black,
                           fill=red!30, inner sep=1pt]
\tikzstyle{arrow}       = [thick,->,>=stealth]

\usepackage{listings}

\definecolor{gray_ulisses}{gray}{0.55}
\definecolor{darkgreen}{rgb}{0.0,0.35,0.0}
\definecolor{keywordblue}{rgb}{0.0,0.0,0.7}
\definecolor{stringred}{rgb}{0.6,0.1,0.1}

\def\codesize{\small}

\lstdefinestyle{cstyle}{
  language=C,
  basicstyle=\ttfamily\codesize,
  keywordstyle=\color{keywordblue},
  commentstyle=\color{gray_ulisses}\itshape,
  stringstyle=\color{stringred},
  numbers=left,
  numberstyle=\footnotesize\color{gray_ulisses},
  numbersep=8pt,
  showstringspaces=false,
  breaklines=true,
  tabsize=4,
  columns=fullflexible,
  keepspaces=true,
  aboveskip=\smallskipamount,
  belowskip=\smallskipamount,
}

\lstnewenvironment{ccode}
{\lstset{style=cstyle,numbers=none}}
{}

\lstnewenvironment{nccode}
{\lstset{style=cstyle}}
{}

\lstdefinelanguage{Coq}{
  morekeywords={Definition,Fixpoint,Lemma,Theorem,Proof,Qed,Defined,forall,
    exists,fun,match,with,end,let,in,if,then,else,Inductive,Record,Notation,
    Ltac,Hypothesis,Variable,Parameter,intros,apply,rewrite,destruct,induction,
    forward,forward_if,forward_call,entailer,Intros,Exists,start_function},
  sensitive=true,
  morecomment=[s]{(*}{*)},
  morestring=[b]",
}

\lstdefinestyle{coqstyle}{
  language=Coq,
  basicstyle=\ttfamily\codesize,
  keywordstyle=\color{keywordblue},
  commentstyle=\color{gray_ulisses}\itshape,
  stringstyle=\color{stringred},
  numbers=none,
  showstringspaces=false,
  breaklines=true,
  tabsize=2,
  columns=fullflexible,
  keepspaces=true,
  mathescape=true,
  aboveskip=\smallskipamount,
  belowskip=\smallskipamount,
}

\lstnewenvironment{coqcode}
{\lstset{style=coqstyle}}
{}

\lstMakeShortInline[style=cstyle]@

\usepackage{commands}

\hypersetup{
  pdftitle={Agent-Driven Verification of Memory Safety for liblzma Decoder
    Components with VST},
  pdfauthor={Prokhor Shlyakhtun, Alexander Gryzlov, Vladimir Kukharenko,
    Vasilii Nesterov, Nikolai Vasiliev, Kirill Ziborov, Eugene Zolotarev,
    Alex Pokras},
  pdfsubject={Formal verification of liblzma decoder components using VST and
    proof-generating agents},
  pdfkeywords={formal verification, memory safety, Verified Software Toolchain,
    liblzma, proof-generating agents}}

\title{Agent-Driven Verification of Memory Safety\\
       for liblzma Decoder Components with VST}
\titlerunning{Agent-Driven Verification of liblzma Decoder Components}

\author{Prokhor Shlyakhtun\inst{1} \and Alexander Gryzlov\inst{1,2} \and
  Vladimir Kukharenko\inst{1} \and Vasilii Nesterov\inst{1} \and
  Nikolai Vasiliev\inst{1} \and Kirill Ziborov\inst{1} \and
  Eugene Zolotarev\inst{1} \and Alex Pokras\inst{1}}
\authorrunning{P. Shlyakhtun et al.}

\institute{GenProof, USA\\
  \texttt{\char123 orenty7, alexgryzlov, vladimir, vasa, nick,
  kirill, eugene, alex\char125\char64 genproof.com}
\and
  IMDEA Software Institute, Madrid, Spain}

\begin{document}

\maketitle

\setlength{\topsep}{5pt}
\setlength{\itemsep}{3pt}

\begin{abstract}

We report on the verification of memory safety for decoder
components of \liblzma{}, the compression library underlying \xzutils{}: the
LZMA2 state machine, the LZMA1 decoder it controls, the outer decoding path,
and the shared sliding-window dictionary. Built with the Verified Software
Toolchain (VST), machine-checked body theorems establish memory safety and
partial functional correctness.
Across 27 completed body proofs, the largest covers \texttt{lzma\_decode},
whose 338 source lines expand to 1{,}934 lines of C after preprocessing;
its proof comprises 183{,}268 lines of proof script over 775{,}768 lines of
mechanically extracted goal statements.
The verification exposed undefined behavior in raw
LZMA1 zero-input handling, where range-decoder macros add zero to a null pointer
and subtract two null pointers.

Unlike similar work that synthesizes verified code, we verify pre-existing, production-scale C. AI agents complete proof
goals and propose
refinements; humans write and review models and specifications, and approve
semantic changes; the \Rocq{} kernel checks the proof terms.

With agents constructing the proof scripts, the main engineering problems lay
in translating and modeling production C, building a robust harness for driving
\Rocq{}, and providing feedback for proving agents. VST's assertion logic
expressed every contract required by the development. We describe the pipeline,
coordination mechanisms, and proof-engineering techniques that resolved these
frictions.

\end{abstract}

\section{Introduction}
\label{sec:intro}
\liblzma{}, the compression library underlying \xzutils{}, is a widely used
open-source dependency.
The 2024 \xz{} backdoor (CVE-2024-3094)~\cite{cve-2024-3094} highlighted the
trust placed in this codebase. While that attack used release and build
machinery rather than a memory-safety defect, memory-safety defects are the
largest single class of exploited vulnerabilities in C and C++ code, on the
order of 70\% of serious issues at organizations such as
Microsoft~\cite{miller2019bluehat}. We therefore address the library's own code:
whether decoder bodies can be proved safe under explicit function contracts.

The Verified Software Toolchain (VST)~\cite{appel2014vst,cao2018vstfloyd}
supports foundational verification of C programs through separation-logic
proofs over \CompCert{} semantics~\cite{leroy2009compcert}, but the process
remains labor-intensive. In our pipeline, AI agents construct and
repair proofs and may propose specification refinements. Humans approve every
change to a model or function specification before agents propagate it through
dependent proofs; the kernel of \Rocq{} (formerly \textsf{Coq})\footnote{Legacy
tool and project names such as \texttt{coqc} and \textsf{CertiCoq} retain the
former name.} checks the resulting proof terms. The \liblzma{} case study
examines both this division of labor and the proof engineering needed to
sustain it on production C.

The reported decoder scope contains 22 VST body proofs
closed through \texttt{Qed} with no uses of \texttt{admit} or
\texttt{Admitted}, covering
the LZMA2 framing state machine, outer decoding path, sliding-window
dictionary, six supporting LZMA1 functions, and the LZMA1 decoding core. Each
theorem establishes \Clight{} safety and model-relative partial functional
correctness for its function body, assuming the corresponding contract's
precondition and imported contracts. This safety result rules out
invalid memory accesses that the semantics represents as stuck execution
(\S~\ref{sec:bg-body-theorems}); \S~\ref{sec:evaluation} gives the inventory and
contract boundary. The archived artifact contains the \Rocq{}/VST proof
sources, verified \liblzma{} source snapshot, and build
instructions~\cite{liblzma-verification-artifact}.

At 183{,}268 lines of proof script, the body proof for
\texttt{lzma\_decode} is the largest; \S~\ref{sec:evaluation} explains its
scale.
\VerifiableC{} does not support the function's original Duff's
device\footnote{Duff's device places \texttt{case} labels inside a loop body,
allowing a \texttt{switch} to enter the loop at different points.}\ control
flow, so the verified 338-line C function uses an explicit state machine;
after \CompCert{} preprocessing, its body spans 1{,}934 nonblank lines of C. Proof
construction spanned 1{,}595 agent sessions over 70 days
(\S~\ref{sec:method-scale}). Verification also exposed source-level undefined
behavior in the raw LZMA1 zero-input path:
range-decoder macros add zero to a null pointer and subtract two null pointers,
outside C's defined pointer-arithmetic cases~\cite[6.5.7]{iso-c-n3220}
(\S~\ref{sec:evaluation}).

\label{sec:intro-contributions}
This paper makes the following contributions:

\begin{itemize}[nosep]
\item \textbf{An agent--human workflow for C verification}
  that assigns proof construction and repair to agents while humans retain
  authority over models and function contracts
  (\S~\ref{sec:methodology}, \S~\ref{sec:core}).
\item \textbf{A body-level verification of \liblzma{} decoder components}
  with 22 closed VST body proofs
  (\S~\ref{sec:core}, \S~\ref{sec:evaluation}).
\item \textbf{Agent engineering for long-running proof development:}
  structured \Rocq{} interaction, standalone goal extraction, incremental
  checking, multi-agent orchestration, and durable session handoffs
  (\S~\ref{sec:method-scale}).
\item \textbf{Proof engineering for production C and VST:}
  build-aware \Clight{} generation, model testing against executions, and
  source normalization
  (\S~\ref{sec:friction}).
\item \textbf{Agent-driven proof optimization:}
  a mechanically orchestrated agent swarm that profiles and refactors completed
  proofs to reduce checking time (\S~\ref{sec:method-scale}).
\end{itemize}

\section{Background}
\label{sec:background}
Interpreting the result requires a precise account of what a VST body theorem
guarantees and which decoder components it covers.

\subsection{The Verified Software Toolchain}
\label{sec:bg-vst}

Separation logic extends Hoare logic with memory ownership. Its \(P * Q\)
connective requires disjoint footprints, while the frame rule preserves
untouched state during local reasoning~\cite{reynolds2002separation}. These
features support modular verification of pointer-rich decoder structures.

\VerifiableC{} provides VST's program logic and automation for \Clight{},
\CompCert{}'s C intermediate language~\cite{appel2014vst,cao2018vstfloyd}.
\clightgen{} preprocesses and translates C into a \Clight{} abstract syntax tree
embedded in \Rocq{}. A \emph{function specification} (\emph{funspec}) uses
\texttt{WITH} for logical parameters, \texttt{PRE} for arguments and owned
memory, and \texttt{POST} for return values and updated memory. Assertions such
as \dataat{} and \texttt{field\_at} link logical values to C objects and fields.

A \texttt{semax\_body} theorem proves that one function body satisfies its
funspec assuming the contracts in its global environment
(\texttt{Gprog}). VST scripts use forward symbolic-execution
tactics\footnote{In \Rocq{}, tactics are metaprograms that construct fragments
of the proof term.} such as \texttt{start\_function}, \texttt{forward},
\texttt{forward\_if}, and \texttt{forward\_call}; \texttt{entailer!} handles
entailments. The \Rocq{} kernel checks the assembled proof term.

\paragraph{Scope of body theorems.}
\label{sec:bg-body-theorems}
A verification scope names an entry point and enumerates the in-project
implementation functions whose bodies the result is intended to cover; calls
outside that set, such as library routines, are treated as imports.
\S~\ref{sec:eval-scope} gives the scope for this case study and reports which
individual body theorems are \emph{closed}, i.e., proved through \texttt{Qed}
with no uses of \texttt{admit} or \texttt{Admitted}. Under the funspec
precondition and imported contracts, a closed \texttt{semax\_body} theorem
establishes that execution cannot get stuck in the \CompCert{} \Clight{}
operational semantics.
This rules out every run-time error that this semantics represents as stuck
execution, including memory faults; if the function returns normally, its
result satisfies the model-derived postcondition. Thus \texttt{semax\_body}
supplies \Clight{} safety and model-relative partial functional correctness. It
does not by itself establish termination, total correctness, trace refinement,
or compatibility between separately maintained caller and callee contracts.

\subsection{The \liblzma{} Library}
\label{sec:bg-liblzma}

\paragraph{Target selection.}
We initially considered several alternative C codebases, including
\textsf{SQLite}, \textsf{Lua}, \textsf{Brotli}, and others, but their size or
reliance on features outside VST's supported C subset made them unsuitable.
We encountered only a few such features in \liblzma{};
\S~\ref{sec:fix-c-standards} describes the required transformations.

\paragraph{Decoder structure.}
\liblzma{}'s \textsf{LZMA} decoder interleaves an adaptive binary range
coder~\cite{rissanen1979arithmetic,pavlov2002lzma} with an
\textsf{LZ77}-style dictionary~\cite{ziv1977lz77}. The range coder
reconstructs bits that the LZMA state machine interprets as either a literal
byte or a back-reference $(\mathit{distance},\mathit{length})$ into recently
decoded output. That output is retained in a fixed-size circular buffer (the
\emph{sliding window}); a match copies bytes one at a time modulo the buffer
size, allowing its source and destination to overlap when
\(\mathit{length} > \mathit{distance}\). The resulting proof obligations
combine bit-level normalization, overlapping copies, wrap-around indexing,
and pointer arithmetic.

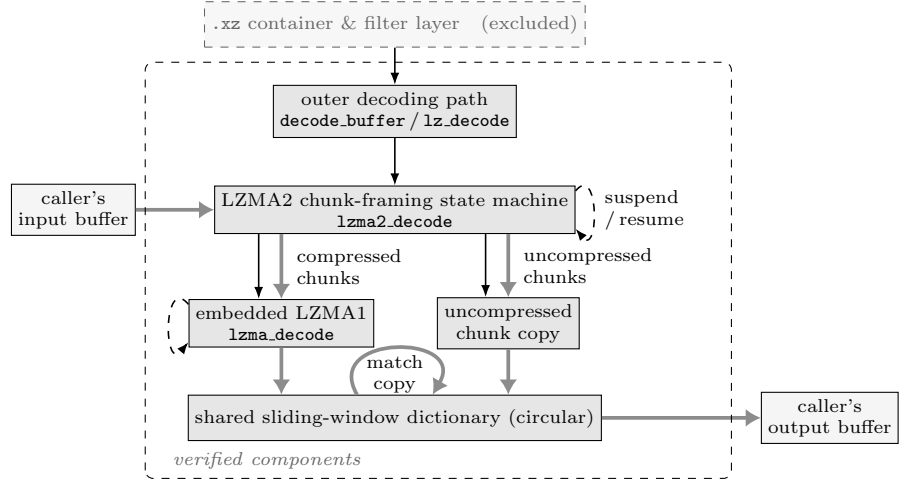
\begin{figure}[t]
  \centering
  \begin{tikzpicture}[
    font=\scriptsize,
    >={Latex[length=1.6mm]},
    component/.style={
      rectangle,
      draw,
      fill=black!10,
      align=center,
      minimum height=0.6cm,
      inner sep=2.5pt
    },
    buffer/.style={
      rectangle,
      draw,
      fill=black!4,
      align=center,
      minimum height=0.6cm,
      inner sep=2.5pt
    },
    excluded/.style={
      rectangle,
      draw=black!60,
      dashed,
      fill=black!3,
      text=black!60,
      align=center,
      minimum height=0.6cm,
      inner sep=2.5pt
    },
    control/.style={->, semithick},
    bytes/.style={-{Latex[length=2mm, width=2.4mm]}, line width=1.4pt, black!45}
  ]

  \node[excluded] (container) at (5.0,1.15)
    {\texttt{.xz} container \& filter layer\quad(excluded)};
  \node[component] (outer) at (5.0,0)
    {outer decoding path\\\texttt{decode\_buffer}\,/\,\texttt{lz\_decode}};
  \node[component] (framing) at (5.0,-1.3)
    {LZMA2 chunk-framing state machine\\
     \texttt{lzma2\_decode}};
  \node[component] (lzma1) at (3.5,-2.8)
    {embedded LZMA1\\\texttt{lzma\_decode}};
  \node[component] (copy) at (6.5,-2.8)
    {uncompressed\\chunk copy};
  \node[component] (dict) at (5.0,-4.05)
    {shared sliding-window dictionary (circular)};
  \node[buffer] (input) at (0.75,-1.3)
    {caller's\\{}input buffer};
  \node[buffer] (output) at (10.75,-4.05)
    {caller's\\output buffer};

  \draw[control] (container) -- (outer);
  \draw[control] (outer) -- (framing);
  \draw[control] ([xshift=-1.8cm]framing.south) --
    ([xshift=-3mm]lzma1.north);
  \draw[control] ([xshift=1.2cm]framing.south) --
    ([xshift=-3mm]copy.north);
  \draw[control, dashed] ([yshift=2.5mm]framing.east)
    to[out=50, in=-50, looseness=2.5]
    node[right=1pt, align=left] {suspend\\ /\,resume} ([yshift=-2.5mm]framing.east);
  \draw[control, dashed] ([yshift=2.5mm]lzma1.west)
    to[out=130, in=-130, looseness=2.7]
    ([yshift=-2.5mm]lzma1.west);

  \draw[bytes] (input) -- (framing);
  \draw[bytes] ([xshift=-1.5cm]framing.south) --
    node[right=2pt, align=left, text=black] {compressed\\chunks} (lzma1.north);
  \draw[bytes] ([xshift=1.5cm]framing.south) --
    node[right=2pt, align=left, text=black] {uncompressed\\chunks} (copy.north);
  \draw[bytes] (lzma1.south) -- (lzma1.south |- dict.north);
  \draw[bytes] (copy.south) -- (copy.south |- dict.north);
  \draw[bytes] ([xshift=-5mm]dict.north)
    to[out=120, in=60, looseness=2.4]
    node[below=1pt, yshift=2.5pt, align=center, text=black] {match\\copy}
    ([xshift=5mm]dict.north);
  \draw[bytes] (dict.east) -- (output.west);

  \draw[dashed, rounded corners=3pt] (1.7,-4.85) rectangle (9.45,0.65);
  \node[anchor=south west, text=black!60, font=\scriptsize\itshape]
    at (1.95,-4.83) {verified components};
\end{tikzpicture}
  \caption{The \liblzma{} LZMA2 decoding path. Thin arrows show control, thick
    gray arrows show byte flow, dashed loops mark suspension and resumption,
    and the dashed enclosure marks the verified scope.}
  \label{fig:decoder-arch}
\end{figure}

\paragraph{Scope.}
Figure~\ref{fig:decoder-arch} sketches the decoding path and the verified
scope. The verified portion spans three source files. In
\texttt{lzma2\_\allowbreak{}decoder.c}, it includes the LZMA2 framing state
machine. In \texttt{lz\_\allowbreak{}decoder.c}, it includes the outer decoding
path and shared sliding-window dictionary. In
\texttt{lzma\_\allowbreak{}decoder.c}, it includes six supporting LZMA1 functions and the structured state-machine implementation of the LZMA1 decoding
core. The scope focuses on these decoder components and excludes the
container/stream layer (\texttt{.xz} parsing and integrity checks), the filter
framework, the index/metadata machinery, and all encoders.

\section{The Agent-Driven Pipeline}
\label{sec:methodology}
Two requirements shape the pipeline: translation of build-configured C and
human control over contract meaning.

\subsection{Pipeline Overview}
\label{sec:method-team}
\Model{} and \Spec{} name stages in our workflow rather than VST artifact
classes. A functional \Model{} gives executable \Rocq{} definitions of a
routine's abstract state and transition. For example, the LZMA2 model
represents the current framing phase, dictionary, and remaining input as
\Rocq{} values; \texttt{lzma2\_decode} maps an initial state to a return code
and successor state. A \Spec{} then adds VST representation predicates that
relate those values to C structs and buffers, and a standard VST funspec whose
postcondition requires the C outcome to agree with the model. Proof agents then
discharge the resulting body obligations by symbolic execution and entailment
solving.
Humans review and approve the models and funspecs; a proof failure may produce
a proposed refinement, but it returns through the same review gate rather
than being applied by the proving agent. The \Rocq{} kernel checks the eventual
proof term. This separation lets proof search proceed autonomously while
contract changes remain subject to human review.

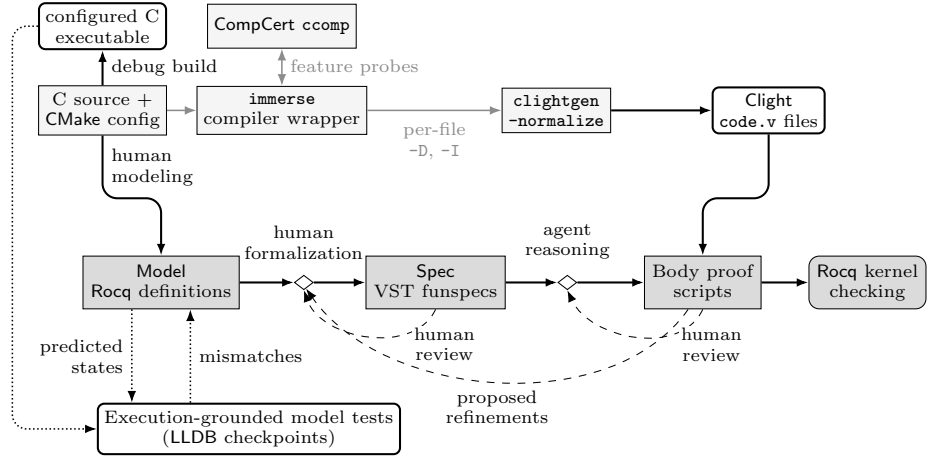
\begin{figure}[t]
  \centering
  \resizebox{\linewidth}{!}{\begin{tikzpicture}[
    font=\scriptsize,
    >={Latex[length=1.6mm]},
    buildstage/.style={
      rectangle,
      draw,
      fill=black!4,
      align=center,
      minimum height=0.6cm,
      inner sep=2.5pt
    },
    artifact/.style={
      rectangle,
      draw,
      rounded corners=2.5pt,
      line width=0.8pt,
      fill=white,
      align=center,
      minimum height=0.6cm,
      inner sep=2.5pt
    },
    stage/.style={
      rectangle,
      draw,
      fill=black!14,
      align=center,
      minimum height=0.72cm,
      inner sep=3pt
    },
    gate/.style={
      diamond,
      draw,
      fill=white,
      aspect=1.3,
      inner sep=1.6pt
    },
    buildflow/.style={->, gray!90, semithick},
    flow/.style={->, thick},
    feedback/.style={->, dashed},
    validation/.style={->, densely dotted, semithick}
  ]

  \node[buildstage] (src) at (0.9,0)
    {C source +\\\textsf{CMake} config};
  \node[buildstage] (wrap) at (3.35,0)
    {\texttt{immerse}\\compiler wrapper};
  \node[buildstage] (ccomp) at (3.35,1.15)
    {\CompCert{} \texttt{ccomp}};
  \node[buildstage] (cg) at (7.1,0)
    {\clightgen{}\\\texttt{-normalize}};
  \node[artifact] (cv) at (10.0,0)
    {\Clight{}\\\texttt{code.v} files};
  \node[artifact] (executable) at (0.9,1.15)
    {configured C\\executable};

  \draw[buildflow] (src) -- (wrap);
  \draw[buildflow, <->] (wrap) --
    node[right] {feature probes} (ccomp);
  \draw[buildflow] (wrap) --
    node[below=0.5mm,align=center] {per-file\\\texttt{-D}, \texttt{-I}} (cg);
  \draw[flow] (cg) -- (cv);
  \draw[flow] (src) --
    node[right] {debug build} (executable);

  \node[stage] (model) at (1.7,-2.35)
    {\Model\\\Rocq{} definitions};
  \node[gate] (g1) at (3.65,-2.35) {};
  \node[stage] (spec) at (5.45,-2.35)
    {\Spec\\VST funspecs};
  \node[gate] (g2) at (7.25,-2.35) {};
  \node[stage] (proof) at (9.1,-2.35)
    {Body proof\\scripts};
  \node[stage, rounded corners] (kernel) at (11.35,-2.35)
    {\Rocq{} kernel\\checking};
  \node[above=1mm of g1,align=center] {human\\formalization};
  \node[above=1mm of g2,align=center] {agent\\reasoning};

  \draw[flow, rounded corners=2mm]
    (src.south) --
    node[right,align=left] {human\\modeling} (0.9,-1.2) --
    (1.7,-1.2) -- (model.north);
  \draw[flow, rounded corners=2mm]
    (cv.south) -- (10.0,-1.2) -- (9.1,-1.2) -- (proof.north);
  \draw[flow] (model) -- (g1);
  \draw[flow] (g1) -- (spec);
  \draw[flow] (spec) -- (g2);
  \draw[flow] (g2) -- (proof);
  \draw[flow] (proof) -- (kernel);

  \draw[feedback] (spec.south)
    to[out=-115,in=-70]
    node[pos=0.2,below=1pt,xshift=4mm,yshift=1mm,
      align=center,fill=white,inner sep=0.5pt]
      {human\\review} (g1.south);
  \draw[feedback] (proof.south)
    to[out=-115,in=-65]
    node[pos=0.2,below=1pt,xshift=4mm,yshift=1mm,
      align=center,fill=white,inner sep=0.5pt]
      {human\\review} (g2.south);
  \draw[feedback] ([xshift=-2mm]proof.south)
    to[out=-125,in=-55,looseness=0.9]
    node[pos=0.5,below,align=center] {proposed\\refinements} (g1.290);

  \node[artifact] (ground) at (2.9,-4.35)
    {Execution-grounded model tests\\(\textsf{LLDB} checkpoints)};
  \draw[validation] ([xshift=-4mm]model.south) --
    node[left,align=right] {predicted\\states}
    ([xshift=-4mm]model.south |- ground.north);
  \draw[validation] ([xshift=4mm]model.south |- ground.north) --
    node[right] {mismatches}
    ([xshift=4mm]model.south);
  \draw[validation, rounded corners=2mm]
    (executable.west) -- ++(-0.35,0) |- (ground.west);
\end{tikzpicture}}
  \caption{The model--specification--proof workflow and C-to-\Clight{}
    ingestion path. Solid black arrows show artifact flow, gray arrows show
    build control, dashed arrows return
    refinements to human review, and the dotted loop connects configured
    executions to model validation.}
  \label{fig:pipeline}
\end{figure}

Figure~\ref{fig:pipeline} shows build-aware \Clight{} generation and model
validation against executions, detailed in
\S~\ref{sec:fix-ingestion} and \S~\ref{sec:fix-grounding}.

Human ownership was organized by artifact: model, specification, and
body-proof contributors reviewed changes to their respective stages against
the C source and adjacent formal artifacts.
The model keeps range decoding and symbol interpretation coupled because
probability updates depend on the LZMA state and dictionary position across
resumable calls.
\S~\ref{sec:core} traces one proposed refinement through this path from
agent report to reviewed, multi-site repair.

\subsection{Coordinating Large Proofs}
\label{sec:method-scale}
\paragraph{The cost of heavy proofs.}
The proofs in this development are unusually expensive to check. A single funspec
proof can run to thousands of lines of forward symbolic execution, and
elaborating one file can take hours and up to 20\,GB of RAM. At that scale the
naive loop of editing a file and re-running \texttt{coqc} is unworkable: every
iteration pays the full cold-elaboration cost from the top of the file. Working
efficiently under these constraints imposes four requirements that together
shaped our tooling. First, iteration must be \emph{interactive}, reusing the
already-checked prefix of a file so that only the tail past an edit is
re-elaborated. Second, the checking backend must stay alive, because every
restart discards the warm cache and forces a cold re-elaboration; restarts must
therefore be rare and narrowly scoped. Third, a backend that runs for a long time
keeps using more and more memory, so it must be able to free memory without
throwing that cache away. Fourth, a proof too large for one file or one agent's
context must be decomposable across files that share a compiled load path. A bare
\texttt{coqc} loop provides none of these. Our \textsf{Claude Code} harness
instead used the Model Context Protocol (MCP)~\cite{mcp2024} to drive
\textsf{rocq-lsp} through \textsf{rocq-mcp}; related MCP-style tooling appears
in~\cite{wang2026}. Runs used \textsf{Claude Code} 2.1.218, under a combination
of a flat-rate subscription and metered API access, on one Ubuntu machine with
128~CPU cores and 1\,TB of RAM, using
\textsf{Coq}~8.20.1\footnote{Version 8.20.1 predates \Rocq{} 9.0, the first
release under the project's new name, and therefore retains the name
\textsf{Coq}.}\ and a project fork of VST with its bundled \textsf{CompCert}
front end.

\paragraph{MCP server.}
We began with LLM4Rocq's \texttt{rocq-mcp}~\cite{llm4rocq-mcp}
(commit \texttt{f9219c8}), the only Rocq MCP server we found. It improved on a
bare \texttt{coqc} command but lacked features these large proofs needed, so we
forked it.
The fork retains eight upstream tools but replaces the
\textsf{Petanque}~\cite{petanque} interactive backend with a client that drives
\texttt{rocq-lsp}~\cite{rocq-lsp} directly; Table~\ref{tab:mcp} summarizes the
resulting differences. In practice the proving agents leaned on four tools:
\texttt{rocq\_get\_state} to inspect the goals at a position;
\texttt{rocq\_query} to search the environment for lemmas and to check types and
definitions; \texttt{rocq\_step} to run a tactic block \emph{speculatively} at a
position and see the resulting goals without modifying the file; and
\texttt{rocq\_compile\_lsp} to check a file incrementally, optionally only
\emph{up to a given position}, so that an expensive or diverging proof further
down never has to run before a lemma near the top can be confirmed. A completed
check can also be marshaled to a \texttt{.vof} snapshot,
\textsf{rocq-lsp}'s serialization of the whole checked document, so a new session
reloads the file's warm cache in seconds instead of re-elaborating it from the
top.

\begin{table}[t]
  \centering
  \caption{Comparison of the two Rocq MCP servers.}
  \label{tab:mcp}
  \begin{tabular}{@{}>{\raggedright\arraybackslash}p{7.0cm}
                      >{\centering\arraybackslash}p{2.0cm}
                      >{\centering\arraybackslash}p{2.4cm}@{}}
    \toprule
    & LLM4Rocq & GenProof (ours) \\
    \midrule
    Interactive backend        & Petanque & rocq-lsp \\
    Gym-like tactic exploration & yes & no \\
    Check up to a point & no & yes \\
    Save \texttt{.vo}/\texttt{.vof} from interactive session & no & yes \\
    Cache trim on high physical-memory use & no & yes \\
    Timeout keeps session warm & no & yes \\
    Per-sentence timing profiling & no & yes \\
    \bottomrule
  \end{tabular}
\end{table}

\paragraph{Extracting and distributing goals.}
Working inside one giant \texttt{semax} body proof imposes three costs. Time: a
helper lemma added near the top of the file is not seen by the main proof until
the whole prefix between them is re-elaborated, so every iteration on it pays
that full elaboration. Memory: while an agent works on one subgoal, Rocq still
holds all the other goals of the proof in memory. Parallelism: a single file is
one editing frontier, so several agents cannot work on different parts of the proof
at once. The natural fix would be to factor the proof into smaller lemmas, but
VST is not built for that: its philosophy is that a function is discharged by a
single \texttt{semax} proof, and a goal arising mid-proof (carrying its full
symbolic-execution context) has no convenient standalone statement to cut it at.
To get the effect anyway, our \texttt{rocq\_extract} tool mechanically cuts a
subgoal out of the live proof: at a chosen point it reads the goal off the
running \textsf{rocq-lsp} session and writes two standalone files next to the
source: the goal as a \texttt{Definition}, its hypotheses lifted into the
statement under universal quantifiers so it closes into a self-contained term,
and a \texttt{Lemma} skeleton whose initial proof state (after re-introducing
those binders) is exactly the state at that point. Because the printed goal is
fully elaborated, with every VST notation expanded to its underlying term, the
generated definition can exceed 100{,}000 lines; agents are told to treat it as
opaque and work only against the \texttt{Lemma} skeleton. The subgoal is then
proved in isolation in its own file, with its own warm cache, by its own agent,
then wired back into the parent with a single \texttt{eapply}. To keep
the detached copy honest, the extraction site is annotated with a
\texttt{confirm\_extraction} marker that hashes the parent goal and detects later
divergence. The marker is a development-only checkpoint: before a body is counted
as closed it is replaced by an ordinary application of the proved lemma, so the
final proof depends on neither the marker nor the extraction plugin, and \Rocq{}
checks the lemma application normally.

\paragraph{Proving agent.}
A proving subagent is scoped to a single proof file and works on that file's open
\texttt{admit}s. It may edit only its assigned file and add proved helper lemmas,
but may never change a definition, a theorem statement, or a funspec. It
\emph{banks} verified work behind explicit, provable \texttt{admit}s so the file
always compiles, and validated progress is never reverted to restore a clean
baseline. A session ends in one of three ways: the assigned goals are closed;
substantial progress is banked and a forward-looking hand-off file is updated for
the next invocation; or the goal is found \emph{unprovable} as stated, in which
case the agent works out why and reports one concrete contract fix: an exact
diff, why it is unavoidable, and why it holds of the C code, \emph{without}
applying it, for human review through the gate in \S~\ref{sec:method-team}.

\paragraph{Orchestration.}
An orchestrator agent manages multi-file proving by spawning proving subagents.
It frees a human from scheduling them and maintains a high-level view of the
proof, since no single subagent holds it whole: a typical session produces a few
hundred new tactics before exhausting its context window, so a large proof is
closed over many hand-off-chained sessions. When a subagent stops, the
orchestrator independently rechecks the file with \texttt{rocq\_compile\_lsp}
(immediate, thanks to the shared warm cache) and commits validated progress. When
a subagent reports a contract fix, the orchestrator drives it down the
human-reviewed path (\S~\ref{sec:method-team}): it applies the change in a
separate worktree, rebuilds, spawns repair subagents for any proofs that break,
and opens a merge request.

\paragraph{Hooks.}
Agents occasionally break the intended workflow out of confusion, so we enforce
the key rules mechanically with \emph{hooks}: scripts the harness runs before or
after a tool call to block it or attach a reminder. Three proved most useful. One
denies reading or editing the ${\sim}100$k-line autogenerated goal files (which
only burn context) while leaving the proof body editable. One keeps verification
on the warm \textsf{rocq-lsp} path by blocking \texttt{coqc} and requiring an
extra confirmation, with a reminder of the workflow, before \texttt{make}, so
agents do not silently fall back to the slow, non-incremental check. One
addresses a recurring misdiagnosis: agents sometimes treated a proof-checking
timeout as a tooling failure and tried to route around it, so a reminder
clarifies that the timeout is a real problem in the proof, which agents then fix,
typically by replacing an over-broad automation tactic with a more targeted one.

\paragraph{Guardrails.}
The \Rocq{} kernel rechecks every accepted proof, and agents may
neither alter trusted funspecs nor introduce admits or axioms into a closed body.
We ran \texttt{Print Assumptions} over every reported body.

\subsection{Optimizing Proof-Checking Time}
\label{sec:fix-agent-tactic}
As with human-written formal proofs, a first complete proof is often not the
fastest to check, and at the volume our pipeline generates, proof \emph{time}
matters as much as proof existence: a tactic that is merely slow in a single hand
proof becomes prohibitive once it runs across hundreds of goals and inside every
agent retry loop. After closing the bodies, we used AI agents to reduce their
checking time. Because the scope was the whole project, we relied on mechanical
orchestration rather than a dedicated orchestrator agent: using the
\textsf{beads} task tracker~\cite{beads} we created one ``optimize file X'' task
per file, and a scheduling script repeatedly spawns an optimizing agent on each
unblocked task; the agent profiles its file, refactors it, and files follow-up
tasks for further ideas, so the swarm manages itself. Most files optimized this
way became at least 30\% faster to check in wall-clock time.

Guided by the per-sentence profiler, the agents concentrated on a few recurring
patterns. The most common was reshaping a goal before one of VST's heavyweight
solvers runs. The grouping of \textsc{prop} conjunctions that \texttt{entailer!}
produces is unstable: adding one conjunct to a loop invariant can fold it into
an unrelated group, and when an existentially quantified stage becomes concrete
\texttt{entailer!} can discharge neighboring conjuncts eagerly. A script written
against one goal shape can therefore fail after a small change. The repair is to
inspect the post-\texttt{entailer!} goal and either script against the observed
shape or use shape-insensitive discharge. Cold-cache time was the other trap: a
tactic that is instantaneous against a warm \textsf{rocq-lsp} cache can exceed a
300-second sentence timeout cold, which an agent misdiagnoses as a logic failure;
the worst offenders were \texttt{entailer!}/\texttt{cancel} over
\texttt{func\_ptr}-laden \textsc{sep} contexts and full reductions over large
composite environments. The countermeasures are targeted: move a handful of
rewrites or reductions ahead of an \texttt{entailer!} or a \texttt{cancel} to
leave the solver with almost nothing to do; extract a single pure fact with
\texttt{sep\_apply} instead of a full \texttt{entailer!}; and never \texttt{simpl}
or \texttt{vm\_compute} a whole model function. Instead, unfold one step and
rewrite under control. These changes repeatedly reduced individual tactic calls
from 40\,s or more to under a second without changing what was proved.

A second recurring optimization avoided pathological elaboration. Reading one
field of a large nested-record representation drove VST's
\texttt{solve\_\allowbreak load\_\allowbreak rule\_\allowbreak evaluation}
tactic, which computes the loaded value, to \texttt{cbv}-reduce the projection
through the entire record type on
every \texttt{forward}, costing 40\,s or more per field access. An agent rebound
this tactic with \textsf{Ltac}'s \texttt{::=} mechanism, without modifying VST or
the specification: for each affected field read the proof establishes, once, a
\texttt{JMeq} equality relating the projection to the loaded value, and the
redefined tactic closes its goal from such an equality when one is present and
otherwise delegates to the original tactic. Each affected \texttt{forward} then
resolves in milliseconds instead of tens of seconds.

\section{A Reviewed Coupling Invariant}
\label{sec:core}
We illustrate the reviewed agent--human interaction through one invariant
repair in the LZMA2 framing decoder proof. The decoder processes a
stream as a sequence of compressed or uncompressed chunks and may return when
an input or output buffer is exhausted, then resume the same chunk on a later
call. Its \texttt{sequence} field records the current phase:
\texttt{Control} reads the next chunk's control byte,
\texttt{Uncompressed1/2} and \texttt{Compressed0/1} parse its size fields,
\texttt{Properties} configures the embedded LZMA1 decoder, \texttt{Lzma}
decodes a compressed payload, and \texttt{Copy} copies an uncompressed one.
During a partially consumed header, \texttt{next\_sequence} records the phase
that follows it.

The VST proof establishes a loop invariant before the first iteration and
re-establishes it after every continuation. The invariant describes the states
that may occur at the loop head and relates the outer LZMA2 phase to the
embedded LZMA1 child's initialization stage. Our functional model records that
stage with four constructors:
\texttt{NoChild} before installation, \texttt{Raw} after installation,
\texttt{Sized} after assigning its output size, and \texttt{DecodeReady} after
reset makes it ready to decode. Treating the outer phase and child stage
independently admits combinations that no execution of the C program can
reach. The coupling clauses exclude those combinations and supply facts needed
to verify the next branch or child-decoder call. The reviewed loop invariant
asserts the following constraints; each is stated with the unreachable state
it excludes.

\begin{enumerate}[label=(\alph*),leftmargin=*,nosep]
\item A \texttt{Raw} child is never at \texttt{Properties} or \texttt{Lzma};
  once properties are available, its next header phase is
  \texttt{Properties}. This excludes using a merely allocated child as though
  its probability state had already been initialized.
\item A \texttt{Sized} child is never in the \texttt{Lzma} or \texttt{Copy}
  sequence. This excludes entering decoding or an uncompressed copy while the
  child is between its output-size assignment and the reset that makes it ready.
\item At \texttt{Control}, any child that is not \texttt{DecodeReady} still
  requires properties. This excludes a control state that advertises completed
  property setup while retaining an incompletely initialized child.
\item During normal header progression, \texttt{next\_sequence} is coupled to
  both the current header and the child stage: \texttt{Copy} follows only a
  compressed-size header with a non-\texttt{Sized} child;
  \texttt{Properties} follows an uncompressed header or a sufficiently
  initialized child; and \texttt{Lzma} requires \texttt{DecodeReady}. This
  excludes stale pairings such as \texttt{Sized}/\texttt{Copy}. The clause is
  conditioned on properties being available, so the immediate error return
  that deliberately leaves \texttt{next\_sequence} stale remains representable.
\end{enumerate}

\paragraph{A reported invariant gap.}
In the \texttt{Compressed1} branch, a proving agent reached an obligation that
required \texttt{next\_sequence <> Copy} when the child was
\texttt{Sized}. The current invariant did not imply it: its \texttt{Copy}
alternative carried no stage constraint and therefore admitted the
\texttt{Sized}/\texttt{Copy} state. Following the frozen-specification rule,
the agent stopped and reported (i) why the goal was not derivable, (ii) why the
combination was nevertheless unreachable in the C code, and (iii) a candidate
strengthening that it did not apply. The source argument was that
\texttt{Sized} arises only on the compressed-chunk path, which sets the next
phase to \texttt{Lzma} or \texttt{Properties}; the uncompressed-copy path that
sets \texttt{Copy} does not resize the child, which remains
\texttt{DecodeReady}.

The specification owner reviewed that reachability argument and approved the
strengthened \texttt{Copy} clause in item~(d). The change was not local: every
producer of the invariant acquired a new obligation. In the next session,
agents re-established the clause at nine producer sites and at its consumers;
a final pass threaded the same fact through the uncompressed-copy branches and
closed the framing body theorem. A condensed version of the report and the
accepted clause appear in Appendix~\ref{app:mr}.
This episode follows the review path in Figure~\ref{fig:pipeline}. We triage a
stuck goal as proof work or a possible specification gap.
Here the agent diagnosed a specification gap, and created a proposal for fixing it together with all affected proofs repaired;
a human approved the strengthened invariant and proposed corrections.

\section{Proof Engineering for Production C and VST}
\label{sec:friction}
The C-to-proof path in Figure~\ref{fig:pipeline} required three engineering
steps: recovering build-configured source, testing the functional model against
executions, and regularizing unsupported C constructs.

\subsection{From Build-Configured C to \Clight}
\label{sec:fix-ingestion}
The pipeline begins with the project's build configuration.
Production C is meaningful only under the exact macros, include paths, and
preprocessor flags its build system selects. Macro expansion itself is not the
problem: \clightgen{} invokes a preprocessor, and a compiler's \texttt{-E}
option can expose the expanded source. Neither mechanism discovers which
definitions and include paths \textsf{CMake} assigns to each translation unit.
Our \texttt{immerse} command runs \textsf{CMake} with a compiler wrapper. During
configuration, \textsf{CMake} compiles small feature-probe programs to detect
compiler and platform capabilities. The wrapper forwards these probes to
\texttt{ccomp}, CompCert's compiler driver, allowing configuration to complete;
during the generated build, it replaces each compilation with \clightgen{}
\texttt{-normalize}, passing that command's \texttt{-D} and \texttt{-I}
options and writing the resulting \Rocq{} file into a mirrored output tree.
Adding a source file to the intercepted build therefore brings it into
\Clight{} generation without a separately maintained list of definitions and
include paths.

\subsection{Grounding the Model in Execution}
\label{sec:fix-grounding}
Execution grounding compares the functional model with concrete runs of the
configured C program before the more expensive body proof. This matters for a
resumable decoder: matching final output may not reveal an incorrect
intermediate range-coder or dictionary state that affects a later call. Human
review still determines model and contract meaning; the comparison provides
complementary evidence of discrepancies.

We use \textsf{LLDB}'s scripting API to place breakpoints at entry to and
return from \texttt{lzma\_decode} and to inspect the buffer and struct states
described by our \dataat{} assertions.\footnote{The technique adapts
\textsf{LLDB}'s script-driven debugging tutorial,
\url{https://lldb.llvm.org/use/tutorials/script-driven-debugging.html}.}
A Python handler reads the live buffers and structs and uses a mapping from
their C layouts to our \Rocq{} representations to lift each call's inputs and
observed outcome. Configurable buffer sizes force both complete calls and
suspension followed by resumption. The snapshots pass through a JSONL trace
into a generated \Rocq{} module; a harness runs the functional model on the
recorded inputs with \texttt{vm\_compute} and compares the model outcome with
the recorded C outcome. This call-level differential test localizes a mismatch
to a particular decoder call and transition.

This process led to six corrective Git commits concerning probability-table
offsets for literals and special distances, the dictionary position used to
compute \texttt{pos\_state}, decoded-literal and repetition tracking,
range-coder state threading through normalization on resumable paths, and
stopping behavior at stream completion and the end-of-payload marker. The tests
exposed these discrepancies before corresponding body-proof construction; the
eventual \texttt{semax\_body} theorem supplies the formal connection between
the C body and its funspec.

\subsection{C Standards Mismatches and Source Normalization}
\label{sec:fix-c-standards}

\paragraph{Pointer arithmetic.}
The \xz{} source deliberately uses index \texttt{-1} for a probability-tree
array:

\begin{nccode}
// -1 is fine, because we start
// decoding at probs[1], not probs[0].
// NOTE: This violates the C standard,
// since we are doing pointer
// arithmetic past the beginning of
// the array.
assert((int32_t)(rep0 - symbol - 1) >= -1);
\end{nccode}

ISO C treats pointer arithmetic outside the same array object, except for the
one-past element, as undefined behavior~\cite[6.5.7]{iso-c-n3220}.
\CompCert{}, however, permits forming out-of-bounds pointers as long as they are
not dereferenced or compared unsafely~\cite[Sec.~6.5.6]{compcert-manual}. In this
case we left the C source unchanged. The VST proof uses the lower-level
\texttt{offset\_val} form rather than
\texttt{field\_address}, \texttt{field\_address0}, or \texttt{ArraySubsc},
which describe in-bounds paths through the structure:

\begin{coqcode}
offset_val (sizeof tushort * offset)
    (field_address dt [StructField _pos_special] ptr)
\end{coqcode}

\paragraph{Clight generation.}
We assigned explicit tags to anonymous structure definitions in the source and
included headers processed by \clightgen{}, giving the generated composite types
stable names without changing their layout or run-time behavior. We also added a
translation-only harness that references the static-inline dictionary
operations so that \clightgen{} emits their bodies; the harness is not called
by the decoder and is outside the verified call graph.

\paragraph{Structure-copying assignment and \texttt{goto}.}
Structure assignment is permitted by ISO C when the left and right operands have
compatible structure or union types~\cite[6.5.17.2]{iso-c-n3220}, and
\CompCert{} supports it. \VerifiableC{}, however, excludes struct-copying
assignments and \texttt{goto} statements~\cite[p.~9]{verifiable-c-manual}. We
therefore normalized the source. Aggregate assignments were expanded into
fieldwise assignments: initialization copies use the macros
\texttt{GENPROOF\_LZMA\_LZ\_DECODER\_INIT} and
\texttt{GENPROOF\_LZMA\_NEXT\_CODER\_INIT}, while the local dictionary and
range-decoder copies are expanded directly. For example:

\begin{nccode}
#define GENPROOF_LZMA_NEXT_CODER_INIT(value) \
	do {                                       \
		lzma_next_coder *ptr = &(value);         \
		ptr->coder = NULL;                       \
		ptr->init = (uintptr_t)(NULL);           \
		ptr->id = LZMA_VLI_UNKNOWN;              \
		ptr->code = NULL;                        \
		ptr->end = NULL;                         \
		ptr->get_progress = NULL;                \
		ptr->get_check = NULL;                   \
		ptr->memconfig = NULL;                   \
		ptr->update = NULL;                      \
		ptr->set_out_limit = NULL;               \
	} while(0)
\end{nccode}

The \texttt{goto} in \texttt{lzma\_decode} jumped to the loop's
\texttt{out} section to save state before returning. It was replaced with a
macro that performs the same state-save block and returns:

\begin{nccode}
#define GENPROOF_GOTO_OUT      \
    /* [...] */                \
    dictptr->pos = dict.pos;   \
    dictptr->full = dict.full; \
    /* [...] */                \
    coder->state = state;      \
    coder->rep0 = rep0;        \
    coder->rep1 = rep1;        \
    coder->rep2 = rep2;        \
    coder->rep3 = rep3;        \
    return ret
\end{nccode}

\paragraph{Duff's device and unstructured \texttt{switch}.}
ISO C permits unstructured \texttt{switch} statements, including Duff's
device. \CompCert{} rejects them by default, but its
\texttt{-funstructured-switch} option accepts them through an optional
front-end source transformation~\cite[pp.~31, 48--49]{compcert-manual}. That
transformation introduces \texttt{goto} labels, so its result remains outside
\VerifiableC{}, which supports neither \texttt{goto} nor unstructured
\texttt{switch}~\cite[p.~9]{verifiable-c-manual}. The main loop of
\texttt{lzma\_decode} is built around Duff's device, so we instead replaced
that control flow with explicit structured state-machine dispatch.

These source changes made the program acceptable to \VerifiableC{}.

\section{Evaluation}
\label{sec:evaluation}
We evaluate three outcomes of the development: the source-level defect it
exposed, the formal scope and scale of the proofs, and the resources used by
the agentic workflow.

\subsection{Undefined Behavior}
\label{sec:eval-ub}
The verification exposed undefined behavior (UB) in the raw LZMA1 zero-input path.

After consuming its five-byte range-decoder prefix, raw LZMA1 decoding permits a
zero-input call: \texttt{lzma\_code} accepts a null \texttt{next\_in} when
\texttt{avail\_in} is zero. The \texttt{rc\_to\_local} macro then adds zero to
the null \texttt{in} pointer.

\begin{nccode}
#define rc_to_local(range_decoder, in_pos, fast_mode_in_required) \
	/* in is a null pointer; in_pos == 0 */ \
	const uint8_t *rc_in_ptr = in + (in_pos); \
	/* ... */
\end{nccode}

Similarly, \texttt{rc\_from\_local} subtracts two null pointer values when it
evaluates \texttt{rc\_in\_ptr - in}. Both operations are UB under the C standard
text used in this development~\cite[6.5.7]{iso-c-n3220}.

\subsection{Verification Scope}
\label{sec:eval-scope}
The reported decoder scope contains 22 closed body theorems for the
decoder components, covering the LZMA2 framing state machine, outer decoding
path, shared sliding-window dictionary, supporting LZMA1 functions, and the
structured state-machine implementation of the LZMA1 decoding core. Each
theorem is proved against its function's funspec. The target excludes the
container/stream layer, filter framework, and all encoders.

The range-decoder operations are inlined into \texttt{lzma\_decode}, so there
are no separate range-decoder C bodies to list. The inlined operations are
supported by five specification lemmas and a standalone well-formedness
development; the latter is not imported into the reported body theorem.

\paragraph{Closed-body inventory and scale.}
Table~\ref{tab:function-sloc} enumerates 27 closed body theorems: 22 within the
decoder scope and five supporting common-layer bodies. It reports function-level
C, specification, and proof SLoC. Here ``closed'' records source status: each
listed body theorem in the audited source snapshot was accepted through
\texttt{Qed} with no uses of \texttt{admit} or \texttt{Admitted}.
At the archived source snapshot~\cite{liblzma-verification-artifact}, the
counted \Rocq{} sources
other than generated \Clight{} ASTs total 1{,}187{,}329 lines: 1{,}042{,}399
code, 120{,}286 comments, and 24{,}644 blank; the generated ASTs
(\texttt{code.v}) add
164{,}102 code lines. The code lines divide into 775{,}768 in generated
\texttt{\_goal.v} files stating extracted proof obligations
(\S~\ref{sec:method-scale}); 183{,}268 in \texttt{lzma\_decode\_body.v} and
the 28 extracted \texttt{ldb\_*\_proof.v} files, plus 39{,}760 in helper
libraries shared across those extracted proofs; 34{,}077 in the remaining body
proofs and module glue; 5{,}314 in
function specifications; 3{,}684 in functional models and the range-decoder
well-formedness development; and 528 in shared infrastructure. The table
attributes proof-file lines to individual functions.

\begin{table}[t]
  \centering
  \caption{Closed body proofs: function-level source lines of code.}
  \label{tab:function-sloc}
  {\scriptsize
  \setlength{\tabcolsep}{2.5pt}
  \renewcommand{\arraystretch}{0.96}
  \begin{tabular}{@{}>{\raggedright\arraybackslash}p{3.5cm}rrr
                      >{\raggedright\arraybackslash}p{5.2cm}@{}}
    \toprule
    Name & Spec & Proof & C & Description \\
    \midrule
    \multicolumn{5}{@{}l}{\textbf{Key functions}} \\
    \texttt{lzma\_decode} & 76 & 183{,}268 & 338
      & Run LZMA1 decoder \\
    \texttt{lzma2\_decode} & 144 & 5{,}643 & 98
      & Run LZMA2 decoder \\
    \addlinespace[2pt]
    \multicolumn{5}{@{}l}{\textbf{Supporting (common layer)}} \\
    \texttt{lzma\_alloc} & 34 & 113 & 13
      & Allocate memory \\
    \texttt{lzma\_alloc\_zero} & 35 & 115 & 16
      & Allocate zero-filled memory \\
    \texttt{lzma\_free} & 33 & 59 & 9
      & Free memory \\
    \texttt{lzma\_bufcpy} & 96 & 392 & 18
      & Copy buffer \\
    \texttt{lzma\_next\_filter\_init} & 42 & 227 & 9
      & Initialize next filter in a chain \\
    \addlinespace[2pt]
    \multicolumn{5}{@{}l}{\textbf{Init chain}} \\
    \texttt{lzma\_lzma2\_decoder\_init} & 323 & 189 & 8
      & Install LZMA2 as the terminal LZ filter \\
    \texttt{lzma2\_decoder\_init} & 148 & 3{,}698 & 23
      & Initialize LZMA2 decoder \\
    \texttt{lzma\_lzma\_decoder\_create} & 117 & 202 & 17
      & Initialize LZMA1 decoder \\
    \addlinespace[2pt]
    \multicolumn{5}{@{}l}{\textbf{Dict}} \\
    \texttt{dict\_get} & 16 & 209 & 7
      & Read byte at distance \\
    \texttt{dict\_get0} & 12 & 43 & 5
      & Read the last byte \\
    \texttt{dict\_is\_distance\_valid} & 14 & 60 & 5
      & Validate distance \\
    \texttt{dict\_repeat} & 31 & 2{,}081 & 47
      & Repeat bytes \\
    \texttt{dict\_put} & 16 & 473 & 7
      & Append byte \\
    \texttt{dict\_put\_safe} & 19 & 115 & 8
      & Check available space and append byte \\
    \texttt{dict\_write} & 55 & 2{,}811 & 13
      & Append bytes \\
    \texttt{dict\_reset} & 11 & 15 & 6
      & Mark dict for reset \\
    \addlinespace[2pt]
    \multicolumn{5}{@{}l}{\textbf{LZ decoder}} \\
    \texttt{lz\_decoder\_reset} & 45 & 88 & 10
      & Reset decoder \\
    \texttt{decode\_buffer} & 56 & 2{,}917 & 39
      & Decode through the dictionary \\
    \texttt{lz\_decode} & 67 & 204 & 45
      & Run LZ decoder \\
    \addlinespace[2pt]
    \multicolumn{5}{@{}l}{\textbf{LZMA1 decoder}} \\
    \texttt{lzma\_decoder\_uncompressed} & 36 & 156 & 8
      & Pre-initialize decoder \\
    \texttt{lzma\_decoder\_reset} & 37 & 3{,}516 & 56
      & Reset decoder \\
    \texttt{lzma\_lzma\_lclppb\_decode} & 28 & 232 & 11
      & Unpack and validate properties \\
    \texttt{literal\_init} & 46 & 148 & 9
      & Reset literal probabilities \\
    \texttt{rc\_read\_init} & 75 & 615 & 15
      & Read the five-byte range-decoder prefix \\
    \addlinespace[2pt]
    \multicolumn{5}{@{}l}{\textbf{LZMA2 decoder}} \\
    \texttt{lzma2\_decoder\_end} & 45 & 415 & 9
      & Free decoder \\
    \bottomrule
  \end{tabular}

  \smallskip
  \raggedright
  Each row specifies source lines of code without empty and comment-only lines:
  its funspec definition, attributed proof file or files, and C function body.
  For \texttt{lzma\_decode}, the proof figure sums
  \texttt{lzma\_decode\_body.v} and the 28 extracted
  \texttt{ldb\_*\_proof.v} files. Shared models and lemmas in other files,
  generated Clight ASTs (\texttt{code.v}), and generated goal-statement files
  (\texttt{\_goal.v}) are not counted in the table.
  \par}
\end{table}

\paragraph{Proof-size accounting.}
Counting only the main lemma in \texttt{lzma\_decode\_body.v} and each of the
28 \texttt{ldb\_*\_proof.v} files gives 101{,}854 code lines; the corresponding
C function has 338 lines. By comparison, the main lemma for the 98-line
\texttt{lzma2\_decode} function needs only 4{,}590 lines. The functions contain
2{,}717 and 120 \Clight{} statement nodes, respectively, yielding 37.5 and 38.3
proof lines per statement, a difference within 2\%. The disparity arises
before verification begins, in
the C-to-\Clight{} expansion (over eight statement nodes per source line
against 1.2): after \CompCert{} preprocessing, the two bodies span 1{,}934 and
110 nonblank lines of C, respectively. The verification-oriented C defines
\texttt{GENPROOF\_GOTO\_OUT}, a macro that saves the local decoder state and
returns when decoding suspends in the middle of a symbol. Preprocessing expands
it 32 times, producing roughly 55 nodes per expansion and about 65\% of the
compiled body. Only 13 invocations occur directly in the source; the rest come
through the \texttt{rc\_*\_safe} range-decoder primitives. For each expansion,
the proof must establish the function's full postcondition from a mid-symbol
state.
\texttt{lzma2\_decode} stays small because it delegates LZMA1 decoding to a
function with a VST specification. Its proof handles that call with one
\texttt{forward\_call}, and its loop invariant groups related memory cells
inside the \texttt{lzma2\_decoder\_stage\_rep} and
\texttt{sliding\_window\_rep} representation predicates. The function suspends
only at the top of its loop. Symbolic execution itself is a minor cost: only
about 2\% of the lines in the 28 extracted proofs are \texttt{forward} steps.
Roughly 42\% of the lines instead belong to blocks repeated verbatim across
multiple proofs. The largest source of repetition is an 18-conjunct
decoder-state well-formedness argument, written out 63 times to re-establish
the loop invariant.
The appendix records the counting method.

\paragraph{Imported contracts and assumptions.}
The closed bodies consume imported contracts. In particular, the LZMA2 module
imports allocator, dictionary, and supporting LZMA1 contracts. The outer module
imports its child-decoder contracts, \texttt{lzma\_bufcpy\_spec}, and VST's
\texttt{memcpy\_spec}. A placeholder funspec
(\texttt{lz\_\allowbreak{}decoder\_\allowbreak{}end\_\allowbreak{}placeholder\_\allowbreak{}spec}) remains in the outer module's
environment.
The project-local assumptions are concentrated in the memory-allocation
interface: the \texttt{malloc} and \texttt{free} body lemmas are declared as
\Rocq{} \texttt{Parameter}s, and an axiom (\texttt{mem\_mgr\_rep}) introduces
the memory-manager predicate. These are the only project-local declarations of
either kind in the proof sources. An audit file runs
\texttt{Print Assumptions} on every reported body theorem; its output lists
these declarations alongside the standard axioms imported with VST.

\subsection{Agentic Development Experience}
\label{sec:eval-agentic}
Project records report 1{,}595 agent sessions over the 70-day
\texttt{lzma\_decode} development interval, using \textsf{Claude Code}
2.1.218 on the machine described in \S~\ref{sec:method-scale}.
Execution-grounded model testing produced six corrective Git commits
(\S~\ref{sec:fix-grounding}), and one reviewed invariant repair was propagated
to nine producer sites (\S~\ref{sec:core}). Finer-grained accounting of session
outcomes, human review, active agent-hours, and service cost would require
recovering and auditing interaction logs that were not available for this
paper.

These figures characterize scale and observed outcomes, not human effort or
labor savings; no human-only baseline is available.

\subsection{Findings}
\label{sec:eval-findings}
For the transformed \texttt{verified-xz} snapshot, the closed bodies establish
\Clight{} no-stuck safety, including memory safety, under their funspec
preconditions and imported contracts. The \texttt{lzma\_decode} precondition
excludes the raw zero-input call of \S~\ref{sec:eval-ub}; applying the result to
upstream \liblzma{} additionally requires source equivalence.

\section{Related Work}
\label{sec:related}
\paragraph{Verifying real C with VST.}
VST provides a foundational separation logic for C whose soundness is tied
to the \CompCert{} semantics, while VST-Floyd supplies the forward symbolic
execution used in our body proofs~\cite{appel2014vst,cao2018vstfloyd}. It has
previously been applied to cryptographic C, including OpenSSL
HMAC~\cite{beringer2015hmac} and SHA-256~\cite{appel2015sha}. We apply the
same foundation to stateful, resumable decoder machinery, placing adaptive
proof search in external agents and keeping model and contract changes under
review.

\paragraph{Agentic LLM experience reports for mechanized proof.}
Three recent reports apply AI agents to substantial \Rocq{} developments.
Paraskevopoulou~\cite{paraskevopoulou2026} uses \textsf{Claude Code} to adapt
an existing \textsf{CPS} proof technique to \textsf{CertiCoq}'s \textsf{ANF}
transformation, producing about 7{,}800 lines of kernel-checked \Rocq{} in 96
hours under high-level human guidance. Wang~\cite{wang2026} uses
\textsf{OpenAI Codex} for supervised, checker-driven repair of
\textsf{CertiGC} proofs invalidated by adding mutation.
Fang and Xiong's software project generation workflow~\cite{fangxiong2026}
uses separate agents to
turn requirements into a verified pure \Rocq{} core, demonstrated by a
generated RISC-V RV32I interpreter, then extracts it to C++/OCaml/Rust. Their
observation that explicit proof goals provide better repair signals than SMT
timeouts echoes our VST experience (\S~\ref{sec:method-scale},
\S~\ref{sec:friction}).

The first two reports concern one deep proof development. We instead verify
multiple functions against human-approved models and specifications.
Their workflow generates new code and leaves effects to an unverified host;
we prove the effects of pre-existing, pointer-heavy C bodies against
\CompCert{} \Clight{} semantics.

\section{Conclusion}
\label{sec:conclusion}
This development closed 22 \texttt{semax\_body} theorems for the listed decoder
components. Under their funspec preconditions and imported contracts, they
establish \Clight{} safety, including memory safety, and model-relative partial
functional correctness; verification also exposed undefined behavior in the raw
LZMA1 zero-input path (\S~\ref{sec:eval-ub}).

Across 1{,}595 \texttt{lzma\_decode} sessions, agents constructed proofs and
propagated human-approved revisions; humans retained authority over models and
specifications, and \Rocq{} checked the resulting terms.

\paragraph*{GenAI Assistance Disclaimer.}
LLM agents assisted proof construction (\S~\ref{sec:methodology}) and manuscript
editing; the authors reviewed all text and retain responsibility for it.

\bibliography{bibliography}

\appendix
\renewcommand*{\theHsection}{appendix.\Alph{section}}
\section{The Proving-Subagent Prompt}
\label{app:prompt}

The prompt below is the actual template our orchestrator instantiated for
every proving subagent it spawned; only the file name, goal list, handoff-file
path, and per-run context vary. It enforces single-file scope: a subagent may
edit exactly one proof file, so parallel agents cannot interfere and every
change is attributable. Definitions, funspecs, and lemma statements remain
immutable; when a goal is unprovable the agent must diagnose why and propose
one concrete fix without applying it. This routes every spec change through
the human review gate of
\S~\ref{sec:method-team} (a worked example is traced in
Appendix~\ref{app:mr}). Completed proof steps are retained while unresolved
goals remain explicit \texttt{admit}s, so the file always compiles, and a
forward-looking handoff file carries state to the next
session; the agent itself cannot commit, and the orchestrator independently
recompiles before it does.

\begin{lstlisting}[basicstyle=\ttfamily\footnotesize,columns=fullflexible,
                   keepspaces=true,breaklines=true,
                   aboveskip=\smallskipamount,belowskip=\smallskipamount]
You are a VST proving subagent for the liblzma-verification repo, spawned in the
background. Read `docs/agent/proving-workflow.md` and follow it. Use the `rocq-mcp`
server -- load its tools with ToolSearch
"select:mcp__rocq-mcp__rocq_get_state,mcp__rocq-mcp__rocq_step,mcp__rocq-mcp__rocq_compile_lsp,mcp__rocq-mcp__rocq_query".

You work ONLY on <FILE>. Goal(s) to push: <admit locations / what to prove>.

FIRST, read <HANDOFF-FILE> if it exists -- it carries the current state, the planned
next steps, and gotchas from the previous run. Keep it updated as you work.

CONTEXT: <orient, don't prescribe -- say what changed and what's now available or
unblocked, not which tactics to run. e.g. "the spec of lzma_decode now has
0 <= effective_size <= Int.modulus; the reach lemma's precondition is now unblocked.">

Reminders (full rules in proving-workflow.md):
- Edit ONLY <FILE>. Never change a definition or a lemma STATEMENT (spec.v,
  *_goal.v, a funspec). If a goal is unprovable, work out WHY and report ONE
  concrete fix -- do NOT apply it.
- Helper lemmas are fine but must be PROVEN.
- Bank verified work with explicit, provable admit.s; NEVER revert validated work.
- You CANNOT commit. When you stop: rocq_compile_lsp the whole file (expect
  errors:[] with any remaining admits) and update <HANDOFF-FILE> (forward-looking:
  state, next steps, gotchas -- no diary).
\end{lstlisting}

\section{An Agent-Reported Specification Gap}
\label{app:mr}

The report-only rule above produces a characteristic artifact. The agent must
explain why the goal is not derivable from the current invariant, justify why
the desired fact nonetheless holds of the C code, and leave a concrete fix for
the owner to apply. The example below is condensed from the internal handoff log of the
framing decoder body proof; it concerns the coupling invariant
between the LZMA2 sequence machine and the embedded LZMA1 child decoder's
initialization stage (\S~\ref{sec:core}). The proof was blocked in
the \texttt{Compressed1} case: with the child in stage \texttt{Lzma1Sized}
and \texttt{next\_sequence = Copy}, the output state's stage validity
requires \texttt{Copy <> Copy}, which is unprovable. The agent's report, lightly
edited for notation:

\begin{enumerate}[label=(\alph*)]
\item \emph{Why it is not derivable.} The coupling hypothesis with
  \texttt{stage = Lzma1Sized} only bounds the next sequence to
  $\{\texttt{Copy}, \texttt{Properties}\}$: the \texttt{Lzma} disjunct is
  excluded by discrimination, but the \texttt{Copy} disjunct of the coupling
  carries no stage constraint. No other loop-invariant conjunct relates
  \texttt{next\_sequence} to the child stage, so the invariant genuinely
  admits the state \texttt{stage = Lzma1Sized} $\wedge$
  \texttt{next\_sequence = Copy}.
\item \emph{Why it is true of the C code.} A \texttt{Sized} child arises
  only from the compressed-chunk header path through
  \texttt{lzma\_\allowbreak decoder\_\allowbreak uncompressed}. Control bytes
  at least \texttt{0x80} set the next sequence to \texttt{Lzma} or
  \texttt{Properties}. Only control values 1 and 2 select \texttt{Copy}; that
  uncompressed path does not touch the child, which remains
  \texttt{DecodeReady} from the previous chunk. Hence
  a \texttt{Lzma1Sized} stage rules out a \texttt{Copy} next sequence in every
  reachable state, but the invariant does not capture this fact.
\item \emph{The concrete fix} (marked ``do NOT apply without owner
  sign-off; spec is frozen''): strengthen the coupling's \texttt{Copy}
  disjunct to carry a stage constraint, mirroring the \texttt{Lzma}
  disjunct, with the caveat (worked out in the report) that this is a
  multi-point strengthening (it obligates every producer of the
  post-condition), not a one-line edit.
\end{enumerate}

The fix the owner accepted is the strengthened disjunct the report asked
for:

\begin{coqcode}
(* coupling, Copy disjunct: 1 conjunct before, 3 after *)
view next_sequence coder = Copy /\ stage <> Lzma1Sized /\
(view sequence coder = Compressed0 \/ view sequence coder = Compressed1)
\end{coqcode}

The internal record dates the report to 2026-06-28 (session M), when it was
accepted by the specification owner; the next
session (N, same day) re-established the strengthened disjunct at nine
producer sites and every consumer, closing the blocked case. The same
invariant issue then occurred again on the uncompressed-copy-chunk paths
and was reported and resolved with the same report-only process; the final
strengthening was threaded in session O, which closed the framing body theorem
with \texttt{Qed}. At no point did a proof
agent edit the specification itself.

\section{Separation Lenses for Struct Footprints}
\label{app:separation-lenses}

During development we used the paper-local term \emph{separation lens} for a
representation technique that regularized proofs over large structs. VST
otherwise represents the decoder's 29-field struct as one deeply nested
spatial assertion, while most operations access only one or two fields. A
separation lens consists of a per-field spatial predicate, a whole-struct
predicate parameterized by the fields temporarily removed from it, and
focus/unfocus lemmas:

\begin{coqcode}
Lemma rep_focus f absent cs p:
  set_mem f absent = false ->
  lzma1_decoder_rep absent cs p
  |-- field_rep cs p f * lzma1_decoder_rep (f :: absent) cs p.

Lemma rep_unfocus f absent cs p:
  set_mem f absent = false ->
  field_rep cs p f * lzma1_decoder_rep (f :: absent) cs p
  |-- lzma1_decoder_rep absent cs p.
\end{coqcode}

Here \texttt{f} selects a field, \texttt{cs} is the logical decoder state,
\texttt{p} is the C struct pointer, and \texttt{absent} records fields already
separated from the whole-struct predicate. \texttt{rep\_focus} extracts one
field's predicate while preserving the remainder; \texttt{rep\_unfocus}
reverses that operation. A proof can therefore perform a local operation while
keeping all other fields framed as a single predicate. This made evolving proof
goals more regular and sped up intermediate versions of the development. The
final proof does not depend on these lenses; we include them because they were
used during development.

\section{Proof-Size Measurement Notes}
\label{app:proof-size}

The proof-code counts reported in \S~\ref{sec:evaluation} count the nonblank
lines remaining after comments are removed. Nested \Rocq{} comments are
handled correctly, and comment delimiters inside string literals are ignored.
Main-lemma figures measure from each \texttt{Lemma} declaration to its
terminating \texttt{Qed}. The aggregate source inventory separately counts
code, comment, and blank lines.

To count macro-expanded C, we preprocess each translation unit with the same
definitions and include paths used by \clightgen{}, then use \CompCert{}'s
parsed-C output to give the expanded code a stable layout. For each function,
we count nonblank lines from its signature through its closing brace,
excluding header contents. We validate the configuration by confirming that
the regenerated \Clight{} statement counts match the artifact.

For the main decoder comparison, \Clight{} statement counts are constructor
counts (\texttt{Sset}, \texttt{Sassign}, \texttt{Sifthenelse},
\texttt{Sloop}, \texttt{Sbreak}, \texttt{Scall}, \texttt{Sreturn},
\texttt{Sswitch}) over the generated \texttt{code.v} abstract syntax trees.
The macro-expansion count is cross-checked against this AST:
\texttt{f\_lzma\_decode} contains 33 \texttt{Sreturn} nodes and exactly 32
assignments to each of \texttt{state}, \texttt{rep0} through \texttt{rep3},
\texttt{probs}, \texttt{symbol}, \texttt{limit}, \texttt{offset},
\texttt{len}, and \texttt{uncompressed\_size}. Each of the three saved
range-coder fields has 64 assignments: one save and one conditional-reset
assignment in each expansion.

The symbolic-execution numerator is the number of \texttt{forward} sentences;
dividing it by the code-line count of the 28 extracted proofs gives 2\%. The
42\% repeated-block share uses exact block matching across their main lemmas
after whitespace normalization.

One dispatch case deserves a note. The \texttt{Normalize} case was proved
across six extracted files totaling 16{,}957 main-lemma lines, against
10{,}071 for the byte-identical fall-through \texttt{IsMatch} case. The
substantial extracted \texttt{Normalize} proofs use 3.53, 2.18, and 0.61 proof
lines per \Clight{} statement node, compared with 4.29 for \texttt{IsMatch}.
For these per-goal ratios, statement counts use the four dominant
constructors (\texttt{Sset}, \texttt{Sassign}, \texttt{Sifthenelse}, and
\texttt{Sreturn}) in each generated \texttt{\_goal.v}, rather than the whole
function's \texttt{code.v}. The extra lines have two sources. First, the proof
case-splits on
\texttt{might\_\allowbreak{}finish\_\allowbreak{}without\_\allowbreak{}eopm},
which is true when the expected uncompressed size is known and fits in the
remaining output space. Both cases separately verify the same code after this
test. Second, the extracted exit paths each re-establish the full postcondition.
The six extracted \texttt{Normalize} proofs were developed first, while the
loop invariants were still being settled; the 22 remaining extracted proofs
were developed later against the settled invariants, each as a single goal.

\end{document}